\documentclass[aps,prl,twocolumn]{revtex4}
\usepackage[utf8]{inputenc}
\usepackage[T1]{fontenc}
 \usepackage{lmodern}
\usepackage{mathptmx}
\usepackage{amsmath,amssymb}
\usepackage{xcolor}
\usepackage{graphicx}
\usepackage{pdfpages}
\graphicspath{{/}{SM/}}
\usepackage[colorlinks,linkcolor=blue,urlcolor=blue,citecolor=blue]{hyperref}

\begin{document}

\title{Dynamical phase transition in the first-passage 
probability of a Brownian motion}

\author{ B. Besga, F. Faisant, A. Petrosyan, S. Ciliberto}\email[E-mail me at: ]{sergio.ciliberto@ens-lyon.fr}

\affiliation{Univ Lyon, ENS de Lyon, Univ Claude Bernard, CNRS,
Laboratoire de Physique, UMR 5672, F-69342 Lyon, France}

\author{Satya N. Majumdar}\email[E-mail me at: ]{satya.majumdar@universite-paris-saclay.fr}

\affiliation{LPTMS, CNRS, Univ. Paris-Sud, Universit\'e
Paris-Saclay, UMR 8626, 91405 Orsay, France}

\date{\today}

\begin{abstract}

{We study the first-passage time distribution (FPTD) $F(t_f|x_0,L)$ 
for a freely diffusing particle starting at $x_0$ in one dimension, to a 
target located at $L$, averaged over the initial position $x_0$ drawn 
from a normalized distribution $(1/\sigma)\, g(x_0/\sigma)$ of finite 
width $\sigma$. We show the averaged FPTD undergoes a sharp dynamical 
phase transition from a two-peak structure for $b=L/\sigma>b_c$ to a 
single peak structure for $b<b_c$. This transition is generated by the 
competition of two characteristic time scales $\sigma^2/D$ and $L^2/D$, 
where $D$ is the diffusion coefficient. A very good agreement is 
found between  theoretical predictions and experimental results 
obtained with a Brownian bead whose diffusion is initialized 
by an optical trap which determines the initial distribution 
$g(x_0/\sigma)$. We show that this transition is robust: it is present 
for all initial conditions with a finite $\sigma$, in all dimensions, 
and also exists for more general stochastic processes going 
beyond free diffusion.}

\end{abstract}

\maketitle

First-passage properties of stochastic processes are fundamental to
understand many important phenomena in nature and have wide ranging
applications across fields~\cite{Redner_book,FPP_book}. These
include estimating reaction rates in chemical processes~\cite{HTB1990,RUK14},
understanding persistence properties in 
nonequilibrium systems~\cite{SM_review,Persistence_review},
computing efficiencies of search algorithms~\cite{BLMV2011,reset_review},
estimating the statistics of extreme events~\cite{EVS_review}
and records in a time series~\cite{MZ_2008,Louven_review,record_review},
numerous applications in biology~\cite{GM2016,SK2019,GHM2020},  
astrophysics~\cite{Chandra_1943} and computer science~\cite{BF_2005}. 

In particular, the 
first-passage properties of a simple random walk or a Brownian motion have
been widely studied, not only as a simple solvable example, but
due to its plethora of applications. One recent application that
has created much interest is in the context of a random walk or a
Brownian motion subjected {to resetting} to its initial
starting point, either
at random times~\cite{EM_2011,EM2011_2,
EM_2014,KMSS14,MSS2015,NG16} or periodically~\cite{PKE16,BBR16}. 
Repeated resetting to its starting position
of a freely diffusing Brownian particle has two major effects:
(i) it drives the particle to a nonequilibrium steady state
so that its position distribution becomes stationary at long times
(ii) the mean first-passage times to a fixed target becomes finite.
Moreover, an optimal resetting rate was found that makes the
mean first-passage time minimal, thus rendering a diffusive search
an efficient search process via resetting~\cite{EM_2011}. 
This led to an enormous recent activities in the field,
both theoretically~\cite{reset_review} and more recently,
experimentally~\cite{Friedman2020,Besga2020}.

There are however
two ways in which realistic situations differ from the assumptions
used in these theoretical models: (a) it is impossible
to reset the particle to its starting point `instantaneously' as
was assumed in the original models (b)
it is physically impossible to reset the particle exactly
to its starting point. The latter situation
arises in particular in experiments conducted with optical 
tweezers~\cite{Friedman2020,Besga2020}, where a particle is usually
trapped in an external confining
potential (optical trap), typically harmonic. At thermal equilibrium, the
stationary position distribution of the particle is thus
a Gaussian with a finite width $\sigma$ (which depends
on the temperature $T$ and the stiffness of the laser trap {$\kappa$ as $\sigma^2 = k_B T/\kappa$}).
The particle is initially prepared in thermal equilibrium
in the trap and then the trap is switched off and the particle undergoes 
free diffusion during a certain period. After this period, the trap
is again switched on and the particle is allowed to relax to its
thermal equilibrium before the trap is switched off again.
The relaxation to thermal equilibrium where the particle
is driven towards the trap center mimics the `resetting' (which
is thus non-instantaneous).
However, under this mechanism, the particle never goes back exactly
to its starting position, but its new starting position for the subsequent
diffusive phase is effectively chosen from a Gaussian 
distribution with a finite width $\sigma$.  
The case $\sigma=0$ (delta function) would correspond exactly to
resetting to the fixed initial position. But in realistic situations, 
$\sigma$ is always finite. 

Several recent {theoretical} studies have addressed the issue (a) that a physical resetting is always non-instantaneous
and the effect of a finite duration of the resetting period is well 
understood~\cite{Reuveni16,EM18,MPCM19a,BS2020,MBMS20,GPKP21}. {On the experimental side the development of protocols accelerating the dynamics of optically trapped colloids \cite{Martinez2016, Chupeau2018} could lead to a drastic reduction of the resetting period.} However,
equally important is issue (b), i.e,
how a finite width $\sigma$ in the initial position
distribution may affect the first-passage properties of the 
Brownian motion under resetting?
Indeed, in the experiment reported in Ref.~\cite{Besga2020}, the
presence of a finite $\sigma$ was found to alter {\em substantially} 
the mean first-passage time to the target.
\begin{figure}
\centering
\includegraphics[width=0.49\hsize]{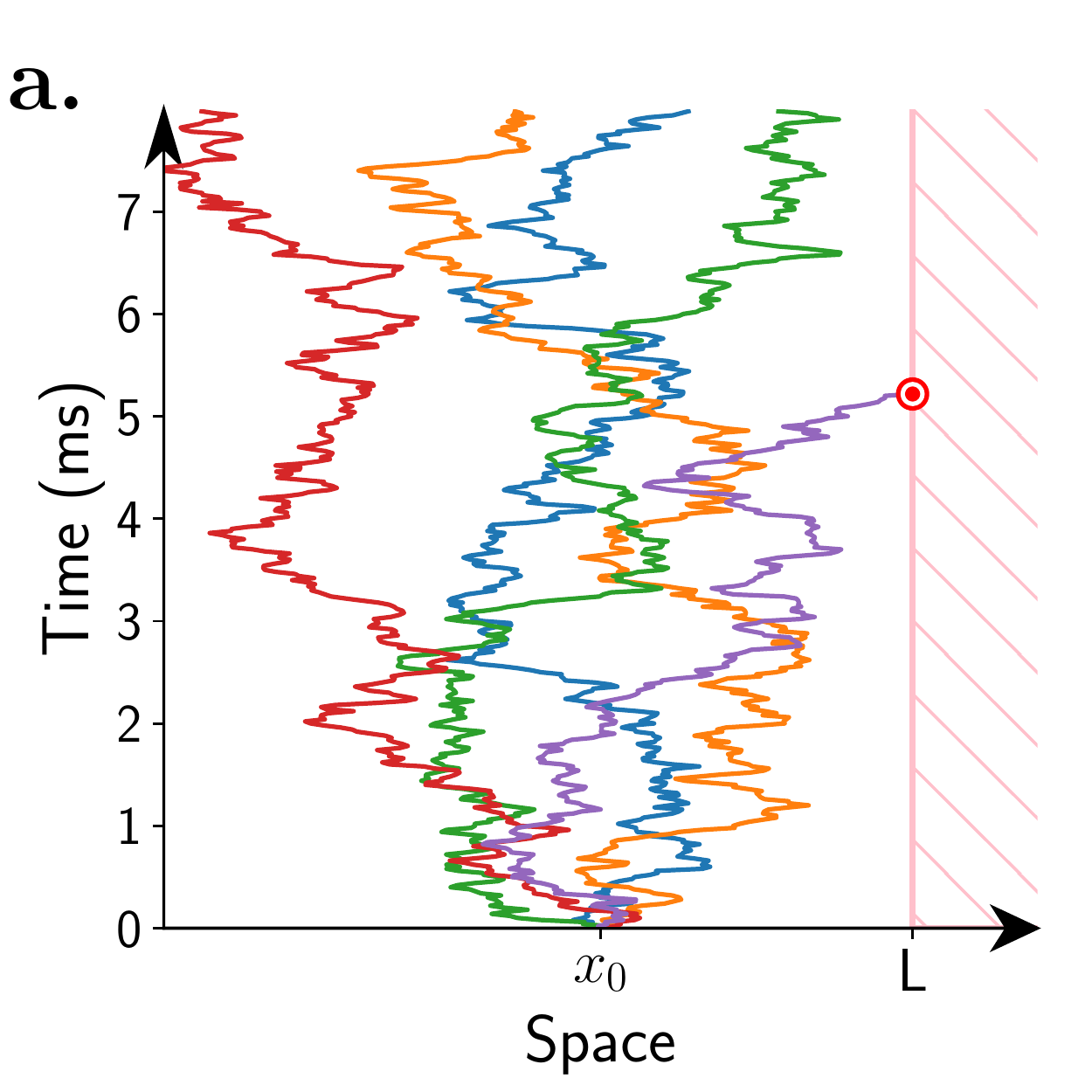} \includegraphics[width=0.5\hsize]{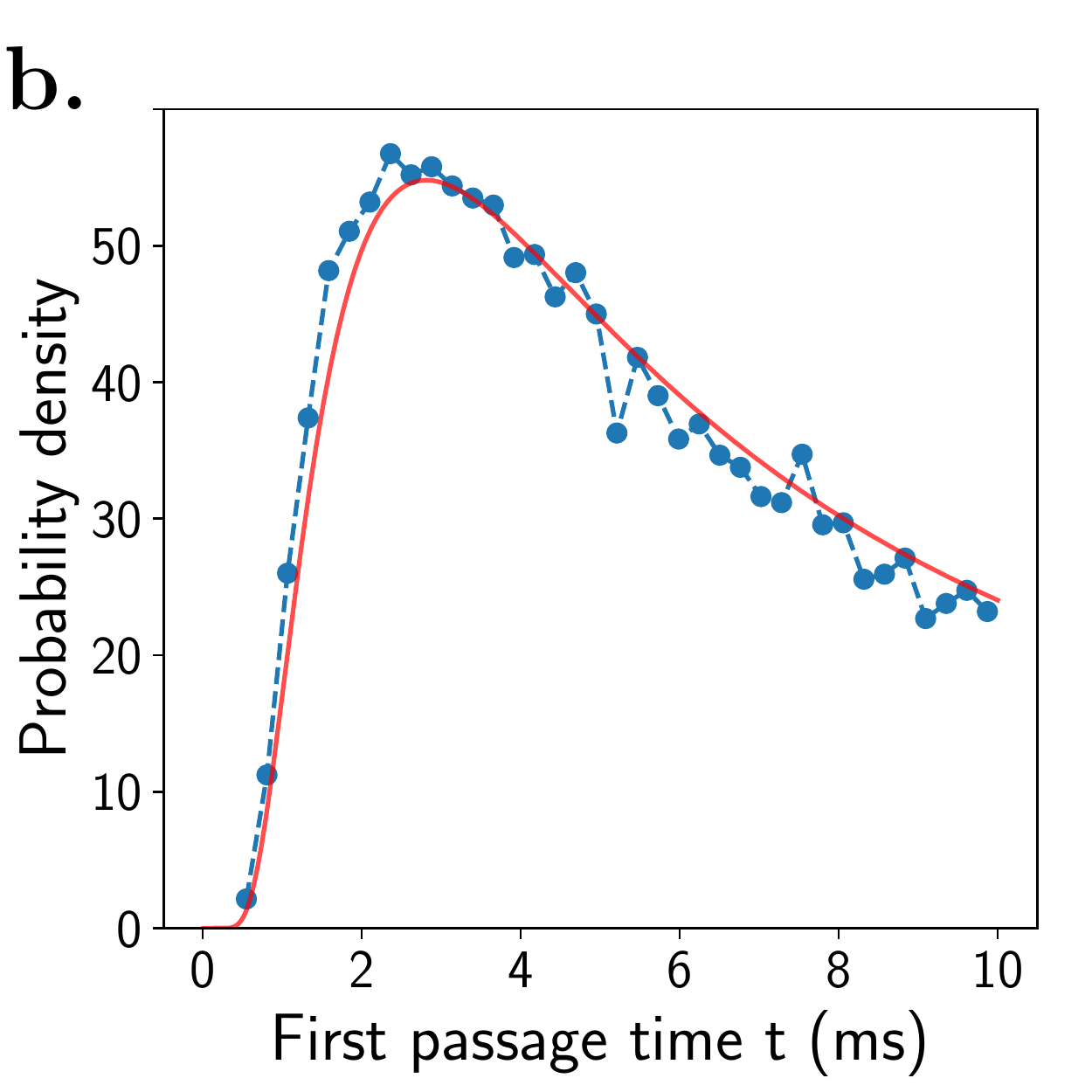} 
\caption{ \textbf{a.}  
Experimental trajectories of a Brownian motion on a line starting at 
$x_0$ with a fixed target at $L$ (parameters : $L - x_0 = 50$ nm and $D 
= 1.5 \times 10^{-13}$ m$^2$.s$^{-1}$). \textbf{b.}  First-passage 
probability density $F(t|0,L)= (L/\sqrt{4\pi Dt^3}) \, e^{-L^2/{4Dt}}$, 
plotted as a function 
of $t$ (red continuous line) for $L = 50$ nm and $D=1.5 \times 10^{-13}$ 
m$^2$.s$^{-1}$ corresponding to the experimental data (blue dots) 
obtained for $1.8 \times 10^4$ first-passages.}
\label{fig:bm}
\end{figure}

The purpose of this Letter is to demonstrate that a finite width $\sigma$ in the initial position distribution
of a Brownian particle profoundly affects its first-passage time
distribution (FPTD), even in the absence of resetting! We start with
an extremely simple system: just a free Brownian motion in
1d {with a diffusion constant $D$,} starting from $x_0$, with
a target located at $L$ (see Fig.\ \ref{fig:bm}a). 
The FPTD $F(t|x_0,L)$
to the target for fixed $x_0$ is well 
known~\cite{Redner_book,Persistence_review}. Now we just average
over the initial position $x_0$ drawn from, say a Gaussian
distribution ${\cal P}(x_0)= e^{-x_0^2/{2\sigma^2}}/\sqrt{2\pi \sigma^2}$
with a finite $\sigma$, {corresponding to the equilibrium
distribution in a harmonic trap before the trap is switched off}. 
We show that just this simple averaging
leads to a profound change in the FPTD.
We have implemented an experiment where we can follow the trajectories 
of a free Brownian particle from an initial gaussian distribution as a 
result of the optical trapping. The experimental setup is detailed in 
\cite{Besga2020} and consists in an infrared laser beam tightly 
focused into a microfluidic chamber to trap a silica micro-sphere of 
radius $R=1$ $\mu$m in water. The position of the Brownian particle is 
read from the deviation of a red laser on a quadrant photodiode at 50 
kHz. The stiffness of the trap is chosen by changing the trapping laser 
intensity power thanks to an electro-optical modulator. In particular 
after equilibration the trap is switched off to follow the free 
diffusion (see Fig.\ \ref{fig:bm}a) and the protocol is repeated to 
acquire statistics on the first passage times.

Our main results are summarised as follows.
A finite $\sigma$ introduces a new time
scale $t_1^*\sim O(\sigma^2/D)$ and for $t\le t_1^*$ the averaged
FPTD develops an anomalous regime. First,
the averaged FPDT diverges as $t^{-1/2}$ as $t\to 0$. 
Secondly, as time increases,
the FPTD decreases, achieves a
minimum at $t\sim t_1^*$, then increases and achieves a maximum
at $t_2^*\sim O(L^2/D)$, before finally
decaying as $t^{-3/2}$ when $t\gg O(L^2/D)$. Thirdly, and most remarkably,
there exists a critical value $\sigma=\sigma_c$ such that
for $\sigma>\sigma_c$, the minimum at $t_1^*$ and the maximum
at $t_2^*$ both disappear and distribution decays monotonically
to $0$ as $t\to \infty$. 
We derive this result analytically, {and demonstrate}
that both the simulation and the experimental data match
perfectly our theoretical predictions. We also provide
a physical meaning of this anomalous regime and the associated
transition at $\sigma=\sigma_c$: we show that for $\sigma<\sigma_c$
when $t_1^*\ll t_2^*$, 
the anomalous early time regime in FPTD
is caused by rare trajectories
that start very close to the target at $L$.
For $t_1^*\ll t_2^*$, such rare atypical trajectories are well separated in time scales
from typical trajectories that start close to the origin. The
transition at $\sigma=\sigma_c$ occurs when these two time scales $t_1^*$
and $t_2^*$ merge with other. In this sense, this phase transition
is `dynamical'.
We then show that this phase transition is robust and
happens for any unbounded initial distribution with a finite width $\sigma$,
not necessarily a Gaussian. Furthermore, we argue and verify numerically
that this transition
is not limited to one dimension, and occurs even in higher dimensions.
Our results are particularly striking since the underlying system
and its associated physics is really very simple.   

We start with a Brownian particle on a line with
diffusion constant $D$, in the presence of a fixed target
at $L$. The particle starts at the initial position $x_0$ (which can be
on either side of $L$).
Let $F(t|x_0,L)\, dt$ denote the probability that the particle
finds the target for the first time in $[t,t+dt]$, given fixed $x_0$ and $L$. 
This FPTD $F(t|x_0,L)$ can be computed
very simply~\cite{Redner_book,Persistence_review}. Let $P(x,x_0,t)$
denote the probability density that the particle reaches $x$ at time $t$,
starting from $x_0$ and does not cross $L$ in time $t$. 
Let us consider $x_0\le L$ (the case $x_0\ge L$
can be similarly computed). 
Then $P(x,x_0,t)$ satisfies the standard Fokker-Planck
equation, $\partial_t P= D\, \partial_x^2 P$, for
$x\le L$ with absorbing boundary
condition at the target $P(x=L,x_0,t)=0$. The solution can
be obtained using the method of images~\cite{Redner_book}
\begin{equation}
P(x,x_0,t)= \frac{1}{\sqrt{4\,\pi\,D\,t}}\left[e^{-(x-x_0)^2/{4Dt}}-
e^{-(x+x_0-2L)^2/{4Dt}}\right]\, .
\label{Fokker_Planck.1}
\end{equation}
Interpreting each trajectory, stating from $x_0$, as an independent Brownian particle,
the FPTD is given simply by the flux of such independent surviving particles 
through $x=L$ at time $t$,
$F(t|x_0,L)= -D\, \partial_x P(x,x_0,t)\Big|_{x=L}$. Using 
\eqref{Fokker_Planck.1}, one gets  
\begin{equation}
F(t|x_0,L)= -\partial_t S(t|x_0,L)= 
\frac{|L-x_0|}{\sqrt{4\,\pi\, D\,t^3}}\, e^{-(L-x_0)^2/{4Dt}}\, .
\label{fp1.1}
\end{equation}

Suppose that $x_0$ is fixed, say at $x_0=0$. Then Eq.\ (\ref{fp1.1}) gives,
$F(t|0,L)= \frac{L}{\sqrt{4\,\pi\, D\,t^3}}\, e^{-L^2/{4Dt}}$.
As a function of $t$ (see Fig. \ref{fig:bm}b) for a plot),
$F(t|0,L)$ has very different behavior across the time scale $t_2^*= L^2/{2D}$.
It decays algebraically as $t^{-3/2}$ for $t\gg t_2^*=L^2/{2D}$. This
is caused by trajectories that start at $0$, but typically diffuse away in the
opposite direction and finally reaches $L$ at times $t\gg t_2^*$. 
In contrast, for $t\ll t_2^*$, it vanishes
extremely rapidly in an essential singular way $\sim e^{-L^2/{4Dt}}$ 
as $t\to 0$. This behavior is also easy to understand physically.
The trajectories that reach $L$ at early times, starting from $0$,
are those that move {\em ballistically} from $0$ to $L$ in time $t$.
Indeed, the statistical weight of a Brownian trajectory is
$\propto \exp[- 1/(4Dt)\, \int_0^t (dx/d\tau)^2\, d\tau]$.
For ballistic trajectories, $dx/d\tau= L/t$ and hence
the flux of such trajectories contribute $\propto e^{-L^2/{4Dt}}$
which exactly reproduces the small $t$ behavior of $F(t|0,L)$. 
Experimentally we compile the first passage times of the Brownian 
particule at a target situated at $L-x_0 = 50$ nm away from its initial 
position and find a very good agreement with the theoretical 
FPTD $F(t|0,L)$ (see Fig. \ref{fig:bm}b).

What happens when the initial position $x_0$ is not fixed, but drawn from a 
distribution ${\cal P}(x_0)$? Averaging
the FPTD $F(t|x_0,L)$ in \eqref{fp1.1} over $x_0$, we get
\begin{equation}
{\overline F}(t|\sigma,L)= \int_{-\infty}^{\infty} F(t|x_0,L)\,
{\cal P}(x_0)\, dx_0\, . 
\label{fpp_av.1}
\end{equation}
For the Gaussian distribution 
${\cal P}(x_0)= e^{-x_0^2/{2\sigma^2}}/\sqrt{2\pi \sigma^2}$,
the integration in \eqref{fpp_av.1} can be performed explicitly.
In terms of the dimensionless variables,
$\tau= \frac{2 D t}{\sigma^2} $ and $b=\frac{L}{\sigma}$,
the averaged FPTD in \eqref{fpp_av.1} can be expressed in the scaling form
\begin{equation}
{\overline F}(t|\sigma,L)= \frac{2D}{\sigma^2}\,
\Phi\left(\frac{2Dt}{\sigma^2}=\tau, \frac{L}{\sigma}=b\right)\, ,
\label{fpp_scaling.1}
\end{equation}
where the scaling function
\begin{eqnarray}
\Phi(\tau,\, b)= \frac{1}{\pi\, \sqrt{\tau}\,(1+\tau)}\, \left[ e^{-b^2/2}+ 
 \sqrt{\frac{\pi\, b^2\, \tau}{2(1+\tau)}}\, 
e^{-b^2/{2(1+\tau)}} \times \right. \nonumber \\
\left.   \times  \,  {\rm erf}\left(\sqrt{\frac{b^2\, \tau}{2(1+\tau)}}\right)\,
\right]\, 
\label{phi_def}
\end{eqnarray}

\begin{figure}
\includegraphics[width=0.9\hsize]{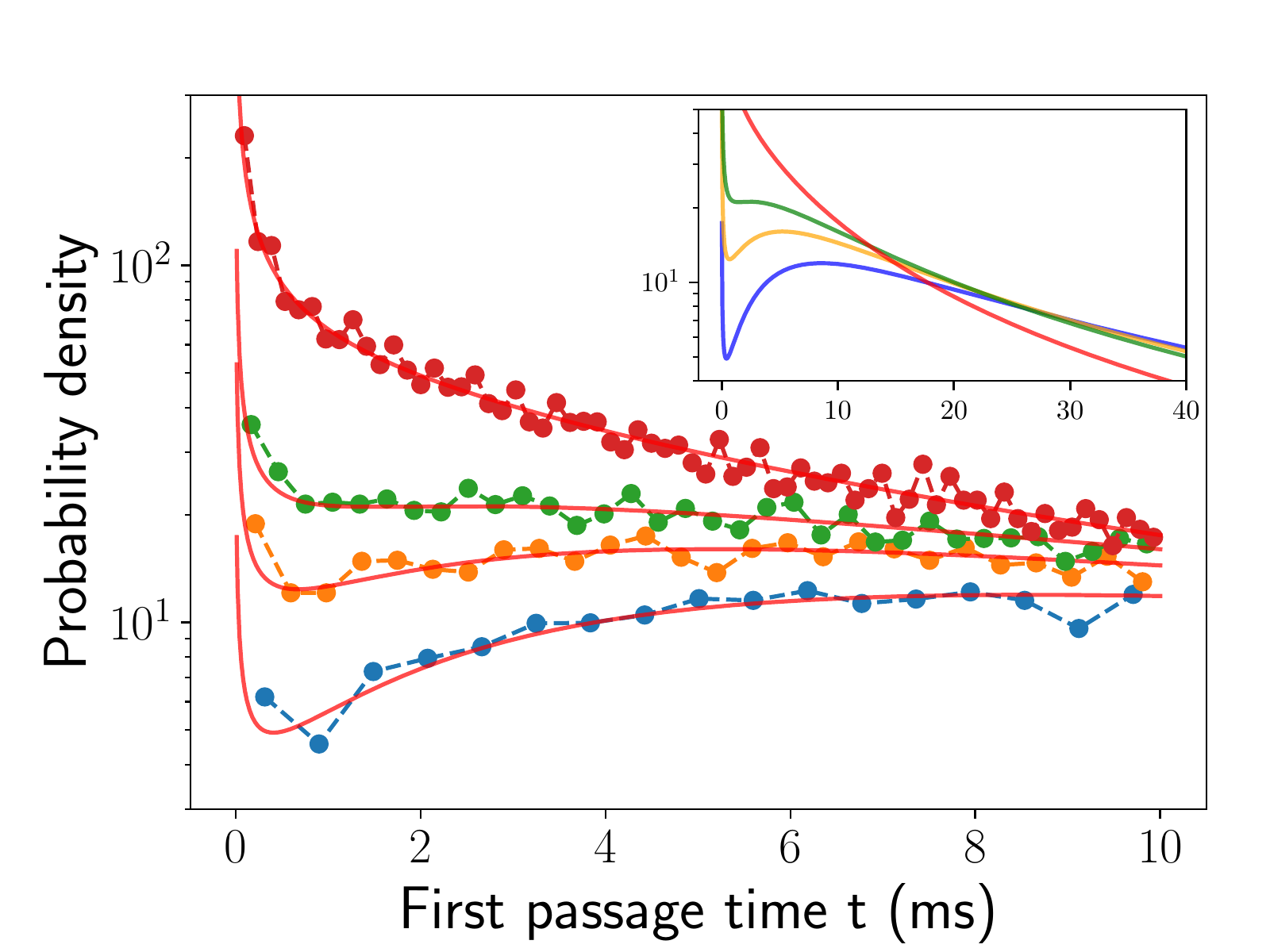}
\caption{\label{fig:fpGauss} 
Theoretical (red lines) an experimental (dots) FPTD 
${\overline F}(t|\sigma,L)$ in \eqref{fpp_scaling.1}, 
for Gaussian initial condition ($\sigma = 36$ nm), 
plotted as a function of $t$ for
$b=3,\, 2.6$, $2.3$ and $1.4$
(from bottom to top) obtained from $4.7 \times 10^3$, $7.9 \times 10^3$, 
$9.1 \times 10^3$  and $1.7 \times 10^4$ first passages time measurements respectively. For any finite $b$
the scaling function diverges as $t^{-1/2}$ as $t \to 0$.
For $b>b_c=2.279$, the function decreases as $t$ increases
and achieves a minimum at $t_1^*\sim O(\sigma^2)$. It then increases
and achieves a maximum at $t_2^*\sim O(L^2)$ and then
decays algebraically as $t ^{-3/2}$ for $t \gg t_2^*$ ({see inset}).
As $b\to b_c$ the two time scales merge and
for $b<b_c$, the function decreases monotonically with
increasing $t$. {Inset : longer timescale for theoretical FPTD as function of t in ms (same parameters than the main figure).}}

\end{figure}

{The average FPTD is plotted in Fig. (\ref{fig:fpGauss}) vs. time} for different 
values of $b$. 
In \eqref{phi_def}, for any fixed $b$, 
the function $\Phi(\tau,b)$ diverges as
$\tau^{-1/2}$ as $\tau\to 0$ (the first term dominates).
Remarkably there is a critical value $b_c=2.279\cdots$ such that
for $b>b_c$, there are two time scales $\tau_1^*\sim O(1)$
and $\tau_2^*\sim O(b^2)$ (in original time $t$ they
correspond to $t_1^*\sim O(\sigma^2/D)$ and $t_2^*\sim O(L^2/D)$
respectively). The FPTD scaling function decreases with increasing
$\tau$ and achieves a minimum at $\tau_1^*$, then increases and achieves 
a maximum at $\tau_2^*$ before finally decaying as $\tau^{-3/2}$ 
for $\tau\gg \tau_2^*$, thus creating a horizontal $S$-shaped
curve {visible in Fig. (\ref{fig:fpGauss}) for ${\overline F}(t|\sigma,L)$}. As $b\to b_c$, the two time scales merge
and for $b>b_c$, the scaling function decays monotonically with
increasing $\tau$. Thus just a simple averaging over
the initial condition leads to a rather rich FPTD, including
a `dynamical' phase transition at $b=b_c$ caused by the merging
of two time scales. The critical value $b_c$ can be precisely
determined as follows. If we plot the derivative $\partial_\tau \Phi(\tau,b)$
as a function $\tau$ or $b>b_c$, it vanishes at the two 
roots $\tau_1^*$ and $\tau_2^*$,
corresponding respectively to the minimum and maximum in 
Fig. (\ref{fig:fpGauss}). As $b\to b_c$, the two roots
approach each other and at $b=b_c$, they merge. Consequently
at $b=b_c$, both the first and the second derivatives
of $\Phi(\tau,b)$ (with respect to $\tau$) vanish at
$\tau_1^*=\tau_2^*=\tau^*$. Solving these two equations (using
{Mathematica software})
for the two unknowms $\tau^*$ and $b_c$, we get
$b_c\approx 2.279$.

Experimentally the Brownian particle is optically trapped in a 
harmonic potential leading to a equilibrium Gaussian distribution in 
the trap with $\sigma = 36$ nm. We release the particle from the trap 
$5 \times 10^{4}$ times and measure the average FPTD for different target 
positions. Our experimental results match very well the theoretical 
predictions (see Fig. (\ref{fig:fpGauss})) and the dynamical phase 
transition associated.

The physics behind this rather striking behavior of the FPTD
can be understood as follows. We can think of the trajectories of the single Brownian
particle as an assembly
of independent Brownian particles with different starting points
$x_0$ and different histories.
Let us first assume that $L\gg \sigma$, i.e., $b\gg 1$
so that the two time scales $\tau_1^*\sim O(1)$ and $\tau_2^*\sim O(b^2)$
are well separated with $\tau_1^*\ll \tau_2^*$. 
We consider the different parts of the
horizontal $S$-shaped FPTD in Fig. (\ref{fig:fpGauss}).
\begin{itemize}

\item Anomalous regime $0\le \tau\le \tau_1^*$: when time $\tau$ is
small, the particles that arrive at $L$ for the first time in $[\tau,\tau+d\tau]$ are the ones that diffuse from the starting points in
the vicinity of the target $L$. Essentially, the particles that
initially are in the region $[L-\sqrt{2Dt}, L+\sqrt{2Dt}]$ will
contribute to this flux at $L$ at time $t$. Thus integrating
Eq.\ (\ref{fpp_av.1}) over this region, it is easy to see that
one gets $\Phi(\tau,b)\sim e^{-b^2/2}/\sqrt{\tau}$. The weight factor
$e^{-b^2/2}= e^{-L^2/{2\sigma^2}}$ is just the probability
of having a particle at $x_0=L$ in the initial condition.
Hence in this regime, the first term in \eqref{phi_def} dominates. {A similar $\tau^{-1/2}$ divergence at short times also
occurs in diffusion controlled reactions with uniform initial
concentration~\cite{Redner_book}.}

\item When $\tau_1^* \le \tau\le \tau_2^*$: As time exceeds 
$\tau_1^*\sim O(1)$, the diffusive particles in the vicinity of 
$L$ have already reached $L$. So, the particles that contribute
to the flux at $L$ at this time are the ones
that start from the center
of the trap $x_0=0$ and reach $L$ ballistically. The weight
of such trajectories $\sim e^{-L^2/{4Dt}}\sim e^{-b^2/{2\tau}}$
makes the minimum around $\tau=\tau_1^*\sim O(1)$. 
In this regime where $1\ll \tau\ll b^2$, the second term
in \eqref{phi_def} dominates.

\item When $\tau\ge \tau_2^*$: For $t\sim L^2/{2D}$, i.e.,
$\tau\sim \tau_2^*\sim O(b^2)$, the
particles that hit $L$ for the first time are the {\em typical} trajectories
that start from the most populated initial region near $x_0=0$
and arrive via diffusion to $L$. Finally when $\tau\gg \tau_2^*$,
the first-passage flux to $L$ are caused by trajectories that start
from the trap center but take much longer times to reach $L$ due
to their sojourns in the direction opposite to $L$.

\end{itemize}

When $b\to b_c$ from above, the two time scales 
$\tau_1^*$ and $\tau_2^*$ merge, the atypical ballistic trajectories 
disappear and the averaged
FPTD is controlled entirely by diffusion --- causing thus a dynamical
phase transition at $b=b_c$ where the two time scales merge.

How robust is this dynamical phase transition? Does it occur for generic
initial conditions, or is it something special for the Gaussian 
case? In fact, consider a generic initial condition
with a finite width $\sigma$ such that, 
${\cal P}(x_0)= (1/\sigma)\, g(x_0/\sigma)$, where $g(y)$ 
is assumed to have an unbounded support
on $y\in [-\infty,\infty]$. Then the integral in \eqref{fpp_av.1} can
still be expressed in the scaling form in \eqref{fpp_scaling.1}, 
with the scaling function given by
\begin{equation}
\Phi(\tau,\, b)=
\frac{\sqrt{2}}{\sqrt{\pi\, \tau}}\, \int_{-\infty}^{\infty} dz\, |z|\,
e^{-z^2}\, g\left(b- \sqrt{2\tau}\, z\right)\, .
\label{sf_gic.1}
\end{equation}
In the limit $\tau\to 0$, we get
\begin{equation}
\Phi(\tau,\, b) \approx
\frac{\sqrt{2}\, g(b)}{\sqrt{\pi\, \tau}}\, \quad {\rm as} \quad
\tau\to 0 \, .
\label{sf_gic.2}
\end{equation}
Thus, for generic unbounded initial condition, the
scaling function diverges {\em universally} as $\tau^{-1/2}$
as $\tau\to 0$. Furthermore, it is not difficult to see that
for any such initial condition $g(y)$, the dynamical transition
at some critical $b_c$ will also exist. As an example, we
consider $g(y)= e^{-|y|}/2$ (double-exponential initial condition),
for which the scaling function can be computed explicitly
\begin{eqnarray}
\Phi(\tau, \, b)= \frac{e^{-b}}{4}\, \left[ \sqrt{\frac{8}{\pi \tau}}
+2\, e^{\tau/2}\, {\rm erf}\left(\sqrt{\frac{\tau}{2}}\right) \right. 
\nonumber \\
\left.  
-e^{\tau/2} \left({\rm erfc}\left(\frac{b-\tau}{\sqrt{2 \tau}}\right)
+ e^{2b}\, {\rm erfc}\left(\frac{b+\tau}{\sqrt{2\tau}}\right)\right)\right]\, .
\label{sf_exp.1}
\end{eqnarray}
When plotted again $\tau$ (not shown here), the scaling function again exhibits
a horizontal $S$-shaped form as in Fig. (\ref{fig:fpGauss}) with a minimum at $\tau_1^*\sim O(1)$
and a maximum at $\tau_2^* \sim O(b^2)$ for $b>b_c\approx2.526$.
The two time scales merge at $b=b_c$ again causing a dynamical phase 
transition. Hence we conclude that this transition is robust and occurs for
any generic unbounded initial condition.

\begin{figure}
	\includegraphics[width=0.8\hsize]{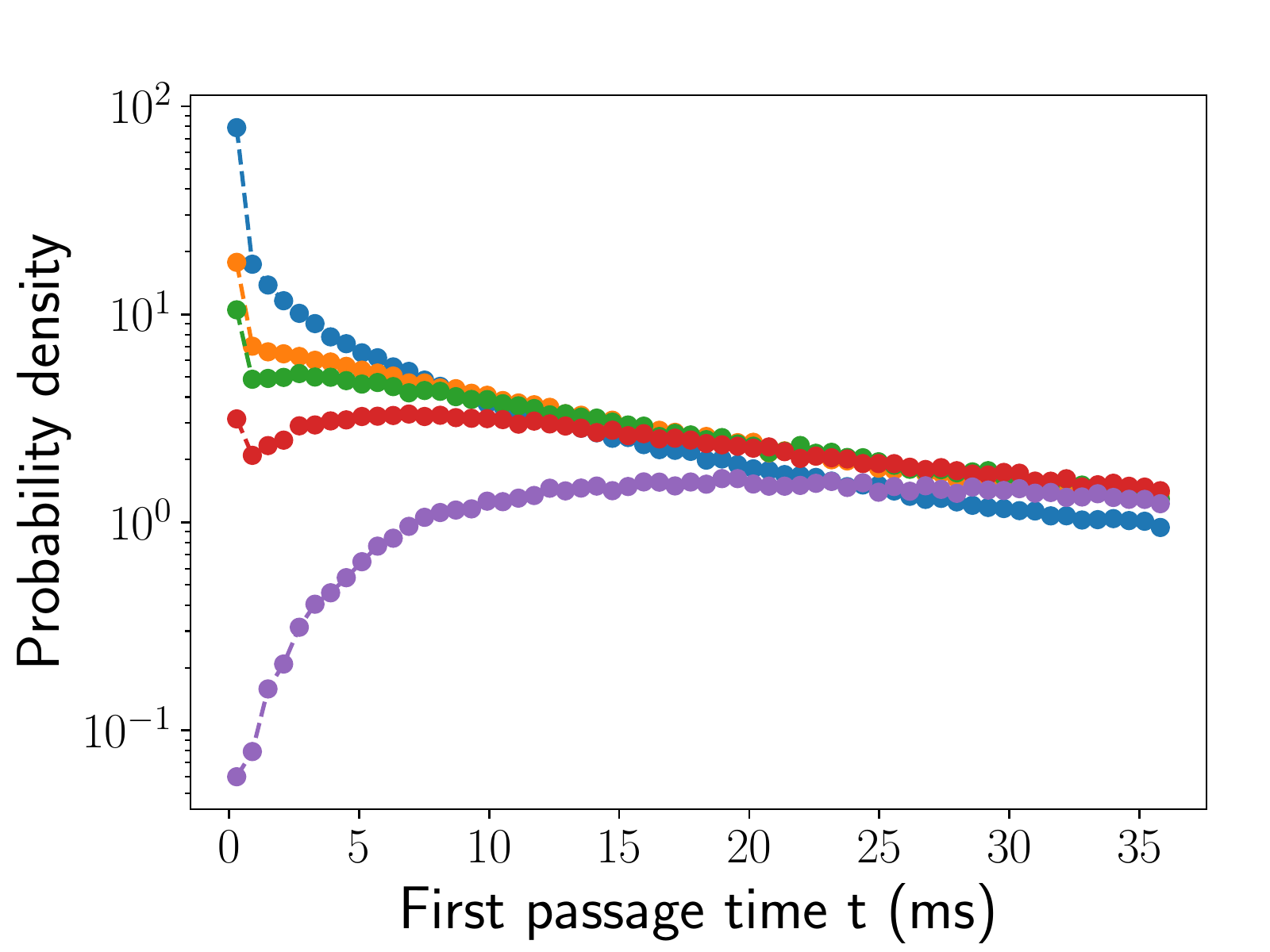}
	\caption{2d numerical simulation of the average FPTD for  
$b = 8, 5.5, 4.5, 4$ and $2$ (from bottom to top) illustrating the dynamical transition 
in 2d around $b_c \approx 4.5$ for $a=0.5$. Plotted for the experimental values $D=1.5 \times10^{-13}$ m$^2$.s$^{-1}$ and $\sigma = 36$ nm.}
	\label{fig:dfp_2_d}
\end{figure}

Is this transition restricted only to one dimension? From the
general physical picture of the problem, it is clear that this
transition should exist even in higher dimensions. In $d>1$, it is
necessary to have a target of a finite `tolerance' size $R_{\rm tol}$, because
Brownian trajectories will surely miss a point target for $d>1$.
Thus we have an additional time scale $t_3^*= R_{\rm tol}^2/{2D}$.
But for fixed $R_{\rm tol}$, the dynamical transition at
some $\sigma_c$ should exist. While the averaged FPTD in $d>1$
can in principle be computed analytically, the calculations are 
somewhat tedious.
However, we have checked numerically that this transition exists for 
$d=2$, by integrating the Langevin equations: 
$\dot x(t)= \sqrt{2\, D}\, \eta_x(t)$ and 
$\dot y(t)= \sqrt{2\, D}\, \eta_y(t)$,
where $\eta_m$ ($m=x,y$) are white noises with
$\langle \eta_m(t)\eta_{m'}(t') \rangle = \delta(t-t') \delta_{mm'}$. 
The simulation starts with $\{x(0),y(0)\}$ 
randomly extracted from a 2d Gaussian distribution of standard 
deviation $\sigma$ and the first passage time is computed when the particle arrives 
within $R_{\rm tol}$ of the target.
We fixed $a=R_\text{tol}/L=0.5$ and simulated $ 5 \times10^6$ trajectories with an integration step of $\approx 50$ $\mu$s. The average FPTD ${\overline F}_{2d}(t|,b,a=0.5)$ is  computed from $\approx 10^6$ first passage times and plotted in Fig.\ \ref{fig:dfp_2_d}. As in the 1d case the average FPTD changes with $b$ 
from a monotonically decreasing function at $b<b_c$ to a 
distribution with a minimum and a maximum for $b>b_c$ with $b_c \approx 4.5$ here.

{Finally, how general is this two-peaked structure
of FPTD and its associated dynamical phase transition? Here we
presented a simple scenario of free diffusion starting from
an initial condition with a finite width $\sigma$, and we have
shown the robustness of this transition with respect to different
initial conditions as well as the spatial
dimension $d$. Does the dynamical transition in FPTD persist for
processes beyond simple diffusion? For example, consider
a particle moving in an external confining potential in 1d, such as
in the Ornstein-Uhlenbeck (OU) process where 
$dx/dt=-\mu\, x +\sqrt{2D}\, \eta(t)$: can one still
see this transition in FPTD? The answer is indeed yes. 
An external confining
potential induces, in addition
to the two time scales $t_1^*\sim O(\sigma^2/D)$ and $t_2^*\sim O(L^2/D)$,
a third time scale 
$\tau_{\rm relax}$ corresponding to the relaxation
time, e.g., in the OU process $\tau_{\rm relax}=1/\mu$. 
As long as $\tau_{\rm relax}\gg {\rm max}(t_1^*,\, t_2^*)$, 
one would still see the competition
between $t_1^*$ and $t_2^*$ and the dynamical transition when they merge.
The algebraic tail of the FPTD for large $t$ corresponding to free diffusion
just gets cut-off by an exponential tail for $t\gg \tau_{\rm relax}$. 
However, it
doesn't affect the short time dynamical phase transition coming
from the interplay between $t_1^*$ and $t_2^*$. We have confirmed this
general picture {with experimental data and} simulations (see the Supp. Mat.~\cite{SM}).
Of course, if the confining potential is very steep $\mu\gg 1$, i.e., 
when $\tau_{\rm relax}$ 
becomes comparable to $t_1^*$ or $t_2^*$, new behaviors of FPTD 
may emerge depending
on the details of the trap. However, the two-peak behavior of FPTD
and its associated dynamical transition is robust as long as the 
confining potential is not too steep.} 

{\acknowledgments
	This work has been partially  supported by the FQXi Foundation, Grant
	No. FQXi-IAF19-05, “Information as a fuel in colloids and
	superconducting quantum circuits.”}


\vfill
\newpage



\section*{Supplementary material}

# Supplementary Material of the Letter: Dynamical phase transition in the first-passage probability of a Brownian motion

B. Besga, F. Faisant, A. Petrosyan, S. Ciliberto*
*Univ Lyon, ENS de Lyon, Univ Claude Bernard, CNRS,*
*Laboratoire de Physique, UMR 5672, F-69342 Lyon, France*

Satya N. Majumdar
*LPTMS, CNRS, Univ. Paris-Sud, Université Paris-Saclay, UMR 8626, 91405 Orsay, France*
(Dated: April 20, 2021)

We report the numerical and experimental results for the probability distribution of the first-passage times (FPTD) for a particle confined in a harmonic potential discussed in the main text of the Letter.

At the end of the main text we discuss the properties of the FPTD when the particle is confined by a potential. The main claim is that in the case of confinement there are three characteristic times. Indeed, in addition to the two time scales $t_1^* \sim O(\sigma^2/D)$ and $t_2^* \sim O(L^2/D)$, there is a third time scale $\tau_{\text{relax}}$ corresponding to the relaxation time inside the potential well. As long as $\tau_{\text{relax}} \gg \max(t_1^*, t_2^*)$, one would still see the competition between $t_1^*$ and $t_2^*$ and the dynamical transition of the FPTD when they merge.

In order to check this claim we study numerically the Brownian motion of a colloid confined in a harmonic potential. Namely we integrate the Langevin equation :

$$\frac{dx}{dt} = -\frac{x}{\tau_{\text{relax}}} + \sqrt{2D}\ \eta(t) \tag{1}$$

where $\eta$ is a delta correlated Gaussian noise with zero mean. The variance of $x$ is $\sigma_x^2 = D\tau_{\text{relax}}$ in the stationary state. We fix $D = 1$ and $t_1 = 1$ which implies that the initial distribution of $x_0$ around $x = 0$ is kept constant with standard deviation $\sigma = 1$. We study the FPTD at $\tau_{\text{relax}}/t_1 = 17.6$ (corresponding to experimental parameters see below) and different values of $t_2 \sim L^2/D$ which is changed by varying the distance to the target $L$, i.e $b = L/\sigma$. The results are compared with those of the free particle in Fig.1. This figure shows that when $\tau_{\text{relax}}$ is larger than $t_1$ and $t_2$ the FPTD is not changed with respect to that of a free particle and the transition occurs at $b \simeq b_c$.

We confirm this observation using the experimental data of a weakly confined Brownian particle for which the trapping laser power was not zero (free diffusion) but was kept very small, getting $\tau_{\text{relax}} = 0.15$ s. As in the

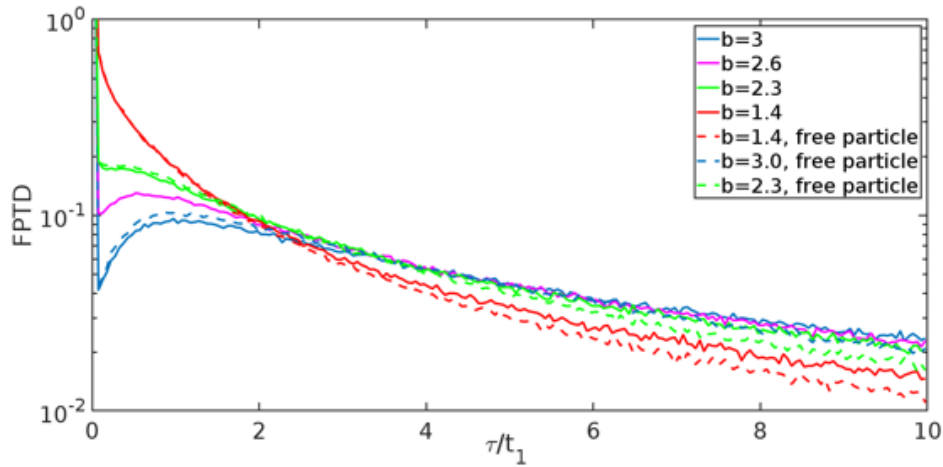

FIG. 1: FPTD (continuous lines) of the OU dynamics defined by Eq. (1) are compared with free diffusing particle (dashed lines) at $t_1 = 1$, $\tau_{\text{relax}}/t_1 = 17.6$ and at the same $b$ of used in fig.2 of the main text.

---

*E-mail me at: sergio.ciliberto@ens-lyon.fr

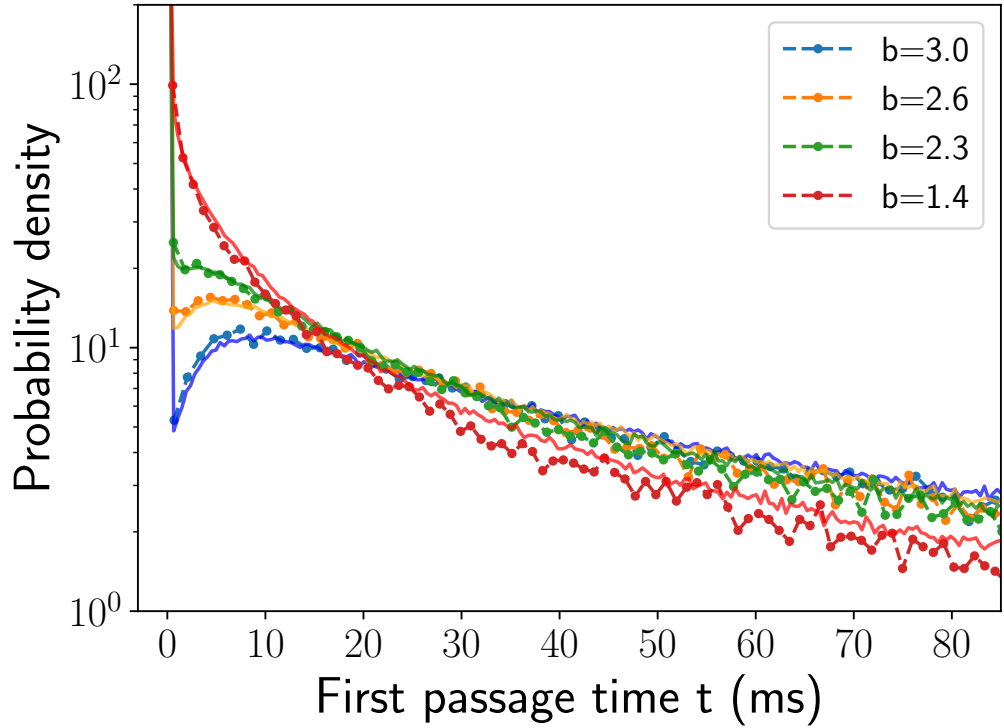

FIG. 2: The experimental data of the FPTD in the case of the OU dynamics are compared with the result of the numerical integration of Eq. (1) at $\tau_{\mathrm{relax}}/t_1 = 17.6$ and various $b$ indicated in the legend.

experiment $t_1 = 8.6$ ms we find that $\tau_{\mathrm{relax}}/t_1 = 17.6$ as for the numerical results of Fig.1. In Fig.2 we compare the results of the numerical simulation of Eq.1 with the experimental values. The agreement is rather good.

These simple examples of FPTD in the motion of a particle confined in a harmonic potential prove that the transition in the FPTD is not a prerogative of free diffusion but it is a general property of the FPTD.

However $\tau_{\mathrm{relax}}$ affects the FPTD when it becomes of the order of $t_2$. We study the FPTD for two values of the ratio $\tau_{\mathrm{relax}}/t_1$ and two values of $b$. The results are compared with those of the free particle in Figs.3 a),b) for two values of $b$. These two figures confirm that as soon as $\tau_{\mathrm{relax}}$ is larger than $t_1$ and $t_2$ the FPTD is very weakly affected with respect to that of a free particle. Instead the effect of the confinement is relevant when $\tau_{\mathrm{relax}}$ is of the order of $t_2$.

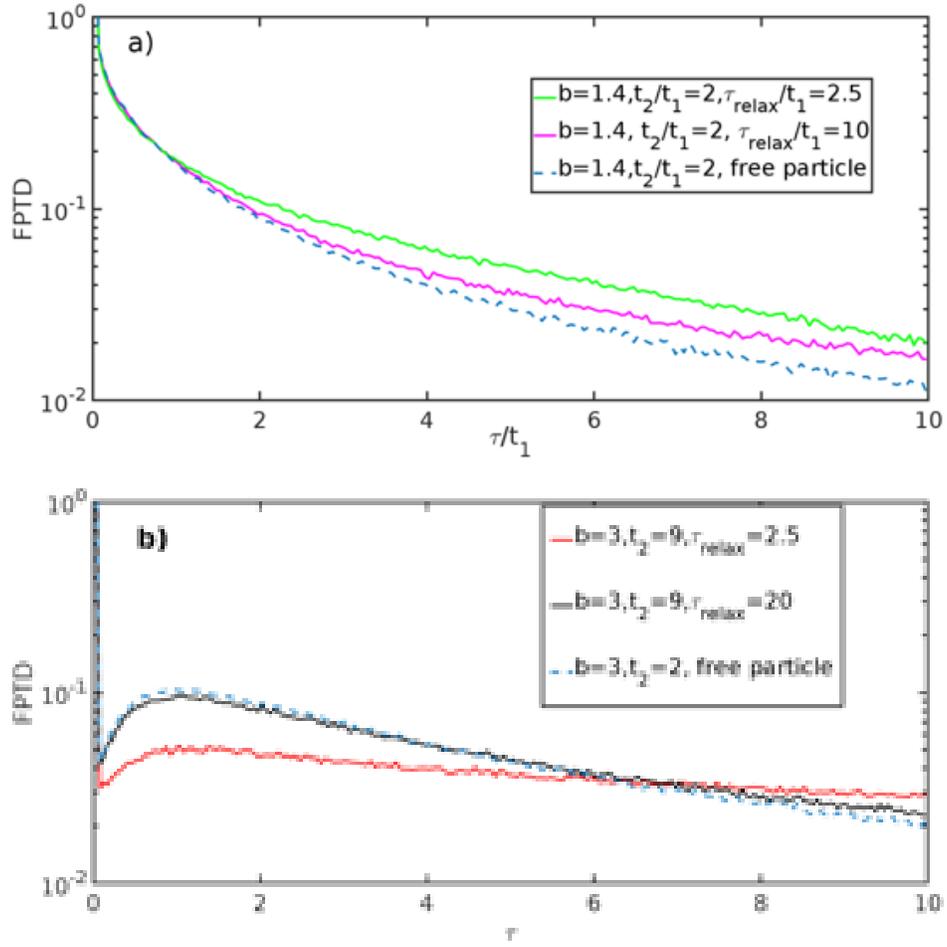

FIG. 3: FPTD of the OU dynamics defined by Eq. (1) are compared with a free diffusing particle at $t_1 = 1$ and various $\tau_{\text{relax}}/t_1$. The other parameters are $b = 1.4, t_2/t_1 = 2$ (a) and $b = 3, t_2/t_1 = 9$ (b)




\begin{thebibliography}{999}

\bibitem{Redner_book} S. Redner, {\em A Guide to First-Passage
Processes} (Cambridge University Press, 2001).

\bibitem{FPP_book} {\em First-Passage Phenomena and Their Applications},
Eds. R. Metzler, G. Oshanin, S. Redner (World Scientific, Singapore, 2013).

\bibitem{HTB1990} P. H\"anggi, 
P. Talkner, and M. Borkovec, {\em Reaction-rate theory: fifty years after Kramers}, Rev. Mod. Phys. {\bf 62}, 251 (1990).

\bibitem{RUK14}
S. Reuveni, M. Urbakh, and J. Klafter, {\em Role of substrate unbinding in Michaelis–Menten enzymatic reactions},
Proc. Natl. Acad. Sci. USA {\bf 111},  4391 (2014).



\bibitem{SM_review} S. N. Majumdar,
{\em Persistence in nonequilibrium systems},
Curr. Sci.  {\bf 77}, 370 (1999).


\bibitem{Persistence_review} A. J. Bray, S. N. Majumdar,
and G. Schehr, {\em Persistence and first-passage
properties in nonequilibrium systems}, Adv. in Phys. {\bf 62}, 225 (2013).

\bibitem{BLMV2011} O. B\'enichou, C. Loverdo, M. Moreau, and R. Voituriez,
{\em Intermittent search strategies}, Rev. Mod. Phys. {\bf 83}, 81 (2011). 

\bibitem{reset_review} M. R. Evans, S. N. Majumdar, and G. Schehr, {\em 
Stochastic resetting and
applications}, J. Phys. A. : Math. Theor. {\bf 53}, 193001 (2020).


\bibitem{EVS_review} S. N. Majumdar, A. Pal, and G. Schehr, {\em 
Extreme value statistics of correlated random variables: A pedagogical
review}, Phys. Rep. {\bf 840}, 1 (2020).

\bibitem{MZ_2008} S. N. Majumdar and R.M. Ziff, {\em 
Universal Record Statistics of Random Walks and L\'evy Flights}, 
Phys. Rev. Lett., {\bf 101}, 050601 (2008).

\bibitem{Louven_review} S. N. Majumdar, {\em Universal first-passage 
properties of discrete-time
random walks and L\'evy
flights on a line:  Statistics of the global maximum and records}, Physica A,
389, 4299 (2010).


\bibitem{record_review} C. Godr\'eche, S. N. Majumdar and G. Schehr, {\em 
Record statistics of a strongly correlated time-series: Random Walks and 
L\'evy Flights}, J. Phys. A: Math. Theor. {\bf 50},  333001 (2017) . 


\bibitem{GM2016}
A. Godec and  R. Metzler, {\em
First passage time distribution in heterogeneity 
controlled kinetics: going beyond the mean first passage time},
Sci. Rep. {\bf 6}, 20349 (2016).


\bibitem{SK2019} J. Shin and A. B. Kolomeisky, 
{\em Target search on DNA by interacting molecules: First-passage approach},
J. Chem. Phys. {\bf 151}, 125101 (2019).

\bibitem{GHM2020} D. S. Grebenkov, D. Holcman, and R. Metzler,
{\em Preface: new trends in first-passage methods and applications 
in the life sciences and engineering}, J. Phys. A: Math. Theor. {\bf 53},
190301 (2020).

\bibitem{Chandra_1943} S. Chandrasekhar, {\em Stochastic problems in 
physics and astronomy},
Rev. Mod. Phys. {\bf 15}, 1 (1943).

\bibitem{BF_2005} S. N. Majumdar, {\em Brownian
Functionals in Physics and Computer Science},
Curr. Sci. {\bf 89}, 2076 (2005).

\bibitem{EM_2011} M. R. Evans and S. N. Majumdar, {\em Diffusion with stochastic resetting},
Phys. Rev. Lett.  {\bf 106}, 160601 (2011).

\bibitem{EM2011_2} M. R. Evans and S. N. Majumdar, {\em Diffusion with optimal resetting},
J. Phys. A: Math. Theor.  {\bf 44}, 435001 (2011).

\bibitem{EM_2014} M. R. Evans and S. N. Majumdar, {\em Diffusion with resetting in arbitrary spatial dimension},
J. Phys. A: Math. Theor. {\bf 47},  285001 (2014).

\bibitem{KMSS14}
L. Ku\'smierz,  S. N. Majumdar,  S. Sabhapandit, and G. Schehr, {\em First order transition for the optimal search 
time of Lévy flights with resetting},
Phys. Rev. Lett. {\bf 113}, 220602 (2014).

\bibitem{MSS2015}
S.N. Majumdar, S. Sabhapandit, and G. Schehr, {\em
Dynamical transition in the temporal relaxation of 
stochastic processes under resetting}, Phys. Rev. E, {\bf 91}, 
052131 (2015).  


\bibitem{NG16}
A. Nagar and S. Gupta S, {\em Diffusion with stochastic resetting at power-law times},
Phys. Rev. E  {\bf 93},  060102 (R) (2016).

\bibitem{PKE16}
A. Pal, A. Kundu, and M. R. Evans, {\em Diffusion under time-dependent resetting},
J. Phys. A: Math. Theor. {\bf 49},  225001 (2016).

\bibitem{BBR16}
U. Bhat, C. De Bacco, and S. Redner, {\em Stochastic search with Poisson and deterministic resetting},
J. Stat. Mech. {\bf 083401} (2016).


\bibitem{Friedman2020} O. Tal-Friedman, A. Pal, A. Sekhon, S. Reuveni, 
and Y. Roichman, {\em Experimental realization of diffusion with stochastic resetting},
J. Phys. Chem. Lett. {\bf 11}, 7350 (2020).

\bibitem{Besga2020} B. Besga, A. Bovon, A. Petrosyan, S. N. Majumdar, and
S. Ciliberto, {\em Optimal mean first-passage time for a Brownian
searcher subjected to resetting: experimental and theoretical results},
Phys. Rev. Res. {\bf 2}, 032029 (2020).

\bibitem{Reuveni16}
S. Reuveni, {\em Optimal stochastic restart renders fluctuations in first passage times universal},
Phys. Rev. Lett.  {\bf 116}, 170601 (2016).

\bibitem{EM18}
M. R. Evans and S. N. Majumdar, {\em Effects of refractory period on stochastic resetting},
J. Phys. A: Math. Theor. {\bf 51} 475003 (2018).


\bibitem{MPCM19a}
A. Mas\'o-Puigdellosas, D. Campos, and V. M\'endez, {\em Transport properties and 
first-arrival statistics of random motion with stochastic reset times}, 
Phys. Rev. E  {\bf 99}, 012141 (2019).

\bibitem{BS2020}
A. S. Bodrova and I. M. Sokolov, {\em Resetting processes with noninstantaneous return}, 
Phys. Rev. E {\bf 101}, 052130 (2020).

\bibitem{MBMS20} G. Mercado-V\'asquez, D. Boyer, S. N. Majumdar, and 
G. Schehr, {\em Intermittent resetting potentials}, J. Stat. Mech.
113203 (2020).


\bibitem{GPKP21} D. Gupta, C. A. Plata, A. Kundu, and A. Pal,
{\em Stochastic resetting with stochastic returns using external trap},
J. Phys. A: Math. Theo. {\bf 54}, 025003 (2021).

\bibitem{Martinez2016} I. A. Mart\'inez, A. Petrosyan, D. Gu\'ery-Odelin, E. Trizac, and S.
Ciliberto, {\em Engineered swift equilibration of a Brownian particle}, Nat. Phys. {\bf 12}, 843 (2016).

\bibitem{Chupeau2018} M. Chupeau, B. Besga, D. Gu\'ery-Odelin, E. Trizac, A. Petrosyan, and S. Ciliberto {\em Thermal bath engineering for swift equilibration}, Phys. Rev. E {\bf98}, 010104(R) (2018)

\bibitem{SM}{Supplemental Material  presents the FPT probability distributions  of the OU processes in a harmonic potential, computed using experimental and numerical data. }


\end{thebibliography}
\end{document}